\documentclass[referee]{aa} 

\usepackage{graphicx}
\usepackage{color}

\usepackage{textcomp}
\DeclareTextSymbolDefault{\degreesymbol}{TS1} 
\DeclareTextSymbol{\degreesymbol}{TS1}{176} 
\DeclareRobustCommand{\textdegree}{\ifmmode\mbox{\degreesymbol}\else\degreesymbol\fi}

\usepackage{lineno}
\usepackage{graphicx}	
\usepackage{amsmath}	
\usepackage{amssymb}	
\usepackage{rotating}
\usepackage{lscape}
\usepackage{comment}
\usepackage[utf8]{inputenc}
\usepackage[dvipsnames]{xcolor}
\usepackage{siunitx}
\usepackage{verbatim}
\usepackage[english]{babel}
\usepackage[T1]{fontenc}
\usepackage{ae,aecompl}
\usepackage{rotating}
\usepackage{textcomp}

\title{Hunting misaligned radio-loud AGN (MAGN) candidates among the uncertain $\gamma$-ray sources of the third Fermi-LAT Catalogue}
\titlerunning{Misaligned AGN}
\author{G. Chiaro\inst{\ref{inst1},\ref{inst2}}\and M. Meyer\inst{\ref{inst5}}\and N.Alvarez Crespo\inst{\ref{inst3}} \and R.J.Britto\inst{\ref{inst4}} \and J.P. Marais\inst{\ref{inst4}}
\\B. van Soelen\inst{\ref{inst4}}\and D. Salvetti\inst{\ref{inst2}}\and G. La Mura\inst{\ref{inst6}}\and D.J Thompson\inst{\ref{inst7}} 
} 
\institute{Corresponding author: Graziano Chiaro, graziani.chiaro@inaf.it \label{inst1}
\and Institute of Space Astrophysics \& Cosmic Physics, INAF, Via Bassini 15, I-20133 Milano Italy\label{inst2}
\and Dept. of Physics, University of Torino, Via P.Giuria Torino, Italy\label{inst3}
\and Dept. of Physics, University of the Free State, Bloemfontein 9300, South Africa.\label{inst4}
\and Kavli Institute for Particle Astrophysics and Cosmology, Dpt. of Physics, SLAC, Stanford University, Stanford, California 94305, USA\label{inst5}
\and Physics \& Astronomy Dept. University of Padova Via Marzolo 8, I-35131 Padova Italy\label{inst6}
\and NASA  Goddard Space Flight Center, Greenbelt, MD 20771 USA\label{inst7}} 
\authorrunning{Chiaro et al.}

\abstract{BL Lac Objects (BL Lacs) and Flat Spectrum Radio Quasars (FSRQs) are radio-loud active galaxies (AGNs) whose 
jets are seen at a small viewing angle (blazars), while Misaligned Active Galactic Nuclei (MAGNs)  are mainly radiogalaxies of type FRI or FRII and Steep Spectrum Radio Quasars (SSRQs),  which show jets of radiation oriented away from the observer's line of sight.  MAGNs are very numerous and well studied in the lower energies of the electromagnetic spectrum but are not commonly observed in the $\gamma$-ray energy range, because their inclination leads to the loss of relativistic boosting of the jet emission.  
The Large Area Telescope (LAT) on board the \emph{Fermi Gamma-ray Space Telescope} in the 100 MeV -- 300 GeV energy range detected only 18 MAGNs (15 radio galaxies and 3 SSRQs) compared to 1144 blazars. 
Studying MAGNs and their environment  in the $\gamma$-ray sky is extremely interesting, because FRI and FRII radio galaxies are respectively considered the parent populations of  BL Lacs and FSRQs, and these account for more than 50~$\%$ of the known $\gamma$-ray sources. The aim of this study is to hunt new $\gamma$-ray MAGN candidates among the remaining blazars of uncertain type  and unassociated AGNs, using machine learning techniques and other physical constraints when strict classifications are not available.
We found 10 new MAGN candidates associated with $\gamma$-ray sources. Their features are  consistent with a source with a misaligned jet of radiation. This study reinforces the need for more systematic investigation of MAGNs in order to improve understanding of the radiation emission mechanisms and and the disparity of detection between more powerful and weaker $\gamma$-ray AGNs.
}

\date{\today}

\begin{document}
\maketitle
\section{Introduction}

Since 2008, the {\it Fermi Gamma-ray Space Telescope} has been observing active galactic nuclei (AGNs), providing an important window into the most extreme phenomenon of the universe:  black holes. The Third {\it Fermi}-LAT Source Catalogue (3FGL) \citep{3fgl} reports $\gamma$-ray data collected in four years of operation (from 2008 August 11 to 2012 July 31) by the {\it Fermi} LAT (Large Area Telescope). The 3FGL Catalogue lists 3033 $\gamma$-ray sources within the 100 MeV--300 GeV energy range, of which 1745 are AGNs, mostly BL Lac objects (BL Lacs) and Flat Spectrum Radio Quasars (FSRQs), the two main kinds of blazars. AGNs with relativistic jets can be classified according  to their jet orientation relative to the line of sight of the observer, as described in the unified model \citep{urry}. In the model, BL Lacs and FSRQs are active galaxies whose jets are seen at a small viewing angle.  Another class of AGNs, Misaligned Active Galactic Nuclei (MAGNs), show a jet of radiation pointed at larger angles to the viewer. The radio-loud MAGNs are separated into two main classes \citep{urry,magn},  namely the edge-darkend Fanaroff-Riley I radio galaxies (FRI) and the edge-brightened Fanaroff-Riley II radio galaxies (FRII) \citep{fan} according to  the distance of the brightest point from the central core.
Steep Spectrum Radio Quasars (SSRQs), whose jet angles are smaller than those of radio galaxies but larger than those of blazars, are often also considered as MAGN \citep{ori}.   We do not include SSRQs in the present analysis. 
There is no exact definition of what \emph{small viewing angle} means. In \citet{sb} a blazar is considered a source with a viewing angle $\theta_ {v}$$\leq$1/{$\Gamma$} where $\Gamma$ is the bulk Lorentz factor of the jet.  
BL Lacs and FSRQs are more easily detected with the {\it Fermi} LAT than MAGNs since the small inclination angle of the jets results in relativistic  boosting of the non-thermal emission.  MAGNs, which are detected in large numbers at radio and optical frequencies, are not so commonly observed in the $\gamma$-ray energy range, because their larger inclinations lead to the loss of this  relativistic boosting.
Only 11 FRI radio galaxies and 4  FRII radio galaxies (Table 1) are classified as MAGNs in the 3FGL Catalogue, compared to 1144 blazars.
One interpretation of this result could  be that {\it Fermi} cannot detect FRIIs because they are far away and therefore not bright enough to be detected. Other possible interpretations are studied by \citet{ele}, but the question remains  open.  More sources are necessary to fully understand this matter.

MAGNs   in the $\gamma$-ray sky remain extremely interesting, because FRI and FRII radio galaxies are respectively considered the parent populations of  BL Lacs and FSRQs \citep{fr}, which account for more than 50~$\%$ of the known $\gamma$-ray sources. In principle, MAGN physics can significantly improve the knowledge of the relationship between the radiation emission mechanisms and the structure of the galaxies, as well as about the disparity of detection between more powerful and weaker $\gamma$-ray sources.  

The aim of this study is to hunt new MAGN candidates among the remaining  blazars of uncertain type (BCUs)  and unassociated AGNs (UCS$_{agn}$s) selected with the machine learning techniques used by \citet{bflap, pablo, zoo}.  In studying MAGNs a great uncertainty  remains about how to compute the jet angle. 
Recent studies by \citet{magn, giov, landt, paola} have introduced some methods for jet angle computing, but the results are still uncertain, and the topic remains  open.  Due to the uncertainty of the methods, we avoided detailed angle calculations and  constrained our hunt to other physical and morphological features of MAGNs and when possible on optical spectroscopy. 
The paper is organized as follows: in Sect. 2 we provide a brief description of the {\it Fermi} LAT, and in Sect. 3 we describe the selection of  MAGN candidates among the uncertain 3FGL objects. In Sect. 4 we present the multifrequency features of the most representative MAGN candidates and their optical spectra. In Sect. 5 we discuss the final results of this study.

\begin{table}

\begin{tabular}{lcccccr}

\hline

\hline

\bf{3FGLname}&\bf{Association name} &\bf{FR type} &\bf{$L_{BLLac}$}\\

\hline

3FGL J0308.6$-$0408&NGC 1218& FRI & 0.80\\

3FGL J0316.6+4119&IC310&FRI & 0.89\\

3FGL J0319.8+4130&NGC1275&FRI & 0.70\\

3FGL J0322.5$-$3712&NCG1316&FRI&0.90\\

3FGL J0334.2+3915&4C+39.12&FRII & 0.26\\

3FGL J0418.5+3013c&3C 111& FRII & 0.04\\

3FGL J0519.2$-$4542&Pictor A&FRII & 0.14\\

3FGL J0627.0$-$3529&PKS 0625-35&FRI & 0.96\\

3FGL J0758.7+3747&NGC 2484& FRI& 0.83\\

3FGL J1145.1+1935&3C 264&FRI&0.81\\

3FGL J1230.9+1224&M87& FRI & 0.90\\

3FGL J1325.4$-$4301& Cen A Core&FRI & 0.70\\

3FGL J1346.6$-$6027& Cen B&FRI & 0.90\\

3FGL J1442.6+5156& 3C 303& FRII & 0.98\\

3FGL J1630.6+8232&NGC 6251& FRI & 0.98\\

\hline

\end{tabular}

\caption{Radio-galaxy MAGNs in the 3FGL {\it Fermi}-LAT Catalogue. }

\end{table}

\section{{\it Fermi} Large Area Telescope}
\label{sec:LAT}

The Large Area Telescope (LAT) is  the primary instrument onboard the {\it Fermi Gamma-ray Space Telescope}, launched by NASA on 2008 June 11 \citep{lat}. It is an imaging, high-energy $\gamma$-ray telescope, covering the energy range from below 100 MeV to more than 300 GeV.  
Operating in survey mode for most of its observations, \emph{Fermi}-LAT has a field of view of 2.4 sr and is able to monitor the whole sky every two orbits (i.e. 2 $\times$ $\sim$96 min). This makes the instrument very appropriate for almost continuous monitoring and search for variability of sources.  The 3FGL Catalogue \citep{3fgl} that forms the basis of the present analysis includes $\gamma$-ray source locations, energy spectra, variability information on monthly time scales, and likely associations with objects seen at other wavelengths. 

\section{The hunting method}
\label{sec:ANN}

Uncertain types of $\gamma$-ray sources often do not have optical spectra available, and without this information a rigorous classification is not possible. Two statistical methods, described by \citet{bflap} and \citet{pablo}, allow a reliable screening by statistical identification of those uncertain sources. The 3FGL Catalogue contains two groups of sources with uncertain classification that offer opportunities to find MAGN candidates: (i) The 559 $\gamma$-ray objects classified as unassociated AGN-like sources (UCS$_{agn}$) in \citep{pablo}, and (ii) the 568 blazar candidates of uncertain type (BCU) studied in \citep{bflap}. These 1127 sources represented the first targets for our search.

We used a machine learning method described by \citet{bflap} based on the Artificial Neural Network (ANN), a  technique that takes advantage of different $\gamma$-ray blazar flaring patterns by applying the Empirical Cumulative Distribution Function (ECDF) to the flux history of the target sources. The output of an ANN can be interpreted as a Bayesian  posterior probability that models the likelihood (\emph{L}) of membership class on the basis of input parameters \citep{gish, rich}. The algorithm computes a likelihood value arranged to have two possibilities: class  \emph{BL Lac} or class \emph{FSRQ}, with a likelihood (\emph{L}) assigned to each analyzed source so that the likelihoods to belong to either of the two classes are related by {$L_{BL Lac}$ = 1 $-$ $L_{FSRQ}$. In this way the closer to 1 is the value of $L_{BL Lac}$, the greater the likelihood that the source is a BL Lac candidate. 
The upper panel of Fig. 1 shows  a testing sample, i.e. a randomly selected 15$\%$ of  3FGL blazars with known classification. The distribution of the likelihood values displays  two distinct and opposite peaks:  BL Lac (blue) and FSRQ (red), the former  at $L_{BL Lac} \sim{1}$ while  the  latter  lies at $L_{BL Lac}\sim{0}$ . 
Since  the testing sample was not used to train the network, the distribution shows the excellent performance of the algorithm in classifying new BL Lac-\emph{like} and FSRQ-\emph{like} candidates.\\
In order to predict the \emph{completeness} of the sample and the fraction of spurious sources labelling  BCUs as BL Lac-\emph{like} or FSRQ-\emph{like} candidates we defined two classification thresholds. The former is based on the optimization of the positive association rate (\emph{precision}), which is defined as the fraction of true positives with respect to the objects classified as positive, of $\sim{90}\%$ , the latter based on  the \emph{sensitivity}, defined as the fraction of objects of a specific class correctly classified as such. 
According to this definition, the BL Lac-\emph{like} classification was characterized by a sensitivity of $\sim{84}\%$,  while  a sensitivity of $\sim{69}\%$ was found for FSRQ-\emph{like} objects.

Since FRI and FRII type radio galaxies are the parent populations of blazar (or blazar-like) sources, we applied to the classified MAGNs listed in the 3FGL Catalogue the optimized algorithm  to recognize blazar-like objects. 
The lower plot in Fig. 1 shows the  known 3FGL radio-galaxy MAGNs likelihood distribution, which resembles blazar-\emph{like} sources.
All of the MAGN lie within the regions previously found for blazars; 12 or 80$\%$ within the BL Lacs (0.6 < $L_{BL Lac}$ < 1.0) region and the remaining 3 or 20$\%$ within the FSRQ likelihood area. This result is consistent with the FRI and FRII classification of 3FGL MAGNs as shown in Table 1, with all FRIs lying at $L_{BL Lac}$>0.6 and all but one FRIIs at $L_{BL Lac}$<0.2. Based on this result we defined two $L_{BL Lac}$ likelihood ranges to hunt MAGN candidates: 0.6<$L_{BL Lac}$<1.0 and 0.0<$L_{BL Lac}$ <0.2. 

Figure 2 shows the likelihood distribution applying the algorithm to the 1127 uncertain sources. The left plot shows the likelihood distribution of the BCUs of the sample; the right plot shows the likelihood distribution of the UCS$_{agn}$s. 
From this analysis, 598 sources are classified as BL Lac-\emph{like} candidates, 180 as FSRQ-\emph{like} candidates, and 349 sources remain unclassified. Considering the BL Lac-\emph{like} + FSRQ-\emph{like} candidates we have a raw sample of 778 objects that include possible MAGN candidates.

From this sample, we identify the most likely MAGN candidates using the properties of known $\gamma$-ray MAGN in the 3FGL Catalogue. The MAGN $\gamma$-ray spectra typically have a power-law index P.I. within a range 1.6 < P.I. < 2.7 \citep{magn,paola}, which is narrower than the range for known blazars, 1.2 < P.I. < 3.1 \citep{3lac}. Furthermore,  known $\gamma$-ray MAGNs have variability indices V.I. that lie between 22.4 and 66.9. This range indicates no significant variability, in contrast to blazars, which are often highly variable (over 40\% of the classified blazars in 3FGL have a V.I. index greater than 72 \citep{3lac}). We therefore applied these same criteria to our sample sources and only retained those  that are within these ranges, leaving 275 candidate objects.\\

Next we turned to multiwavelength information, starting with the radio counterparts of the candidate 3FGL sources, all of which are found in the NRAO VLA Sky Survey (NVSS)\footnote{{\tt http://www.cv.nrao.edu/nvss/}}. We  selected by visual inspection  sources that showed a spatially extended structure, i.e., those that were visibly larger than the 45 arcsec spatial resolution of NVSS.  A more efficient, but more complicated, morphological classification of the candidates could have been done by a survey like the VLA Faint Images of the Radio Sky at Twenty-Centimeters (FIRST) survey \citep{vla}, with angular resolution of about  5 arcsec.  Because the main aim of this study is to verify the process of finding $\gamma$-ray MAGN candidates rather than a rigorous classification of the MAGN sky, we considered the NVSS dataset sufficient for our analysis. In addition, NVSS covers about 3 times as much of the sky as FIRST. This morphological cut eliminates  point-source-like objects more likely to be blazars, whose aligned jets give them a strong core dominance. This final selection leaves 10 candidate sources, listed in Table 2.
This small number of candidates is reasonable, since the ratio of our candidate sample to the initial sample of uncertain sources is comparable to the ratio of $\gamma$-ray MAGNs to classified blazars in the 3FGL Catalogue.  

\begin{figure}
\begin{center}
\includegraphics[width=0.4 \textwidth]{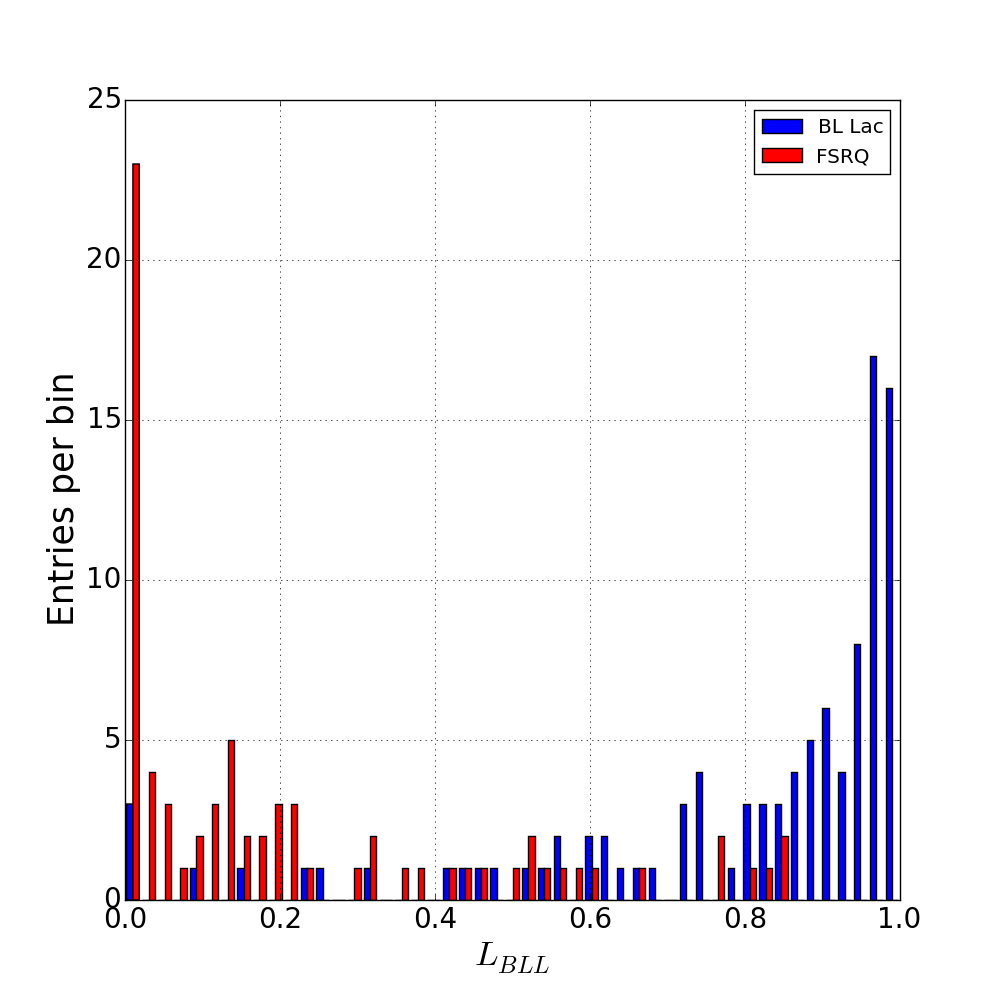}
\includegraphics[width=0.4 \textwidth]{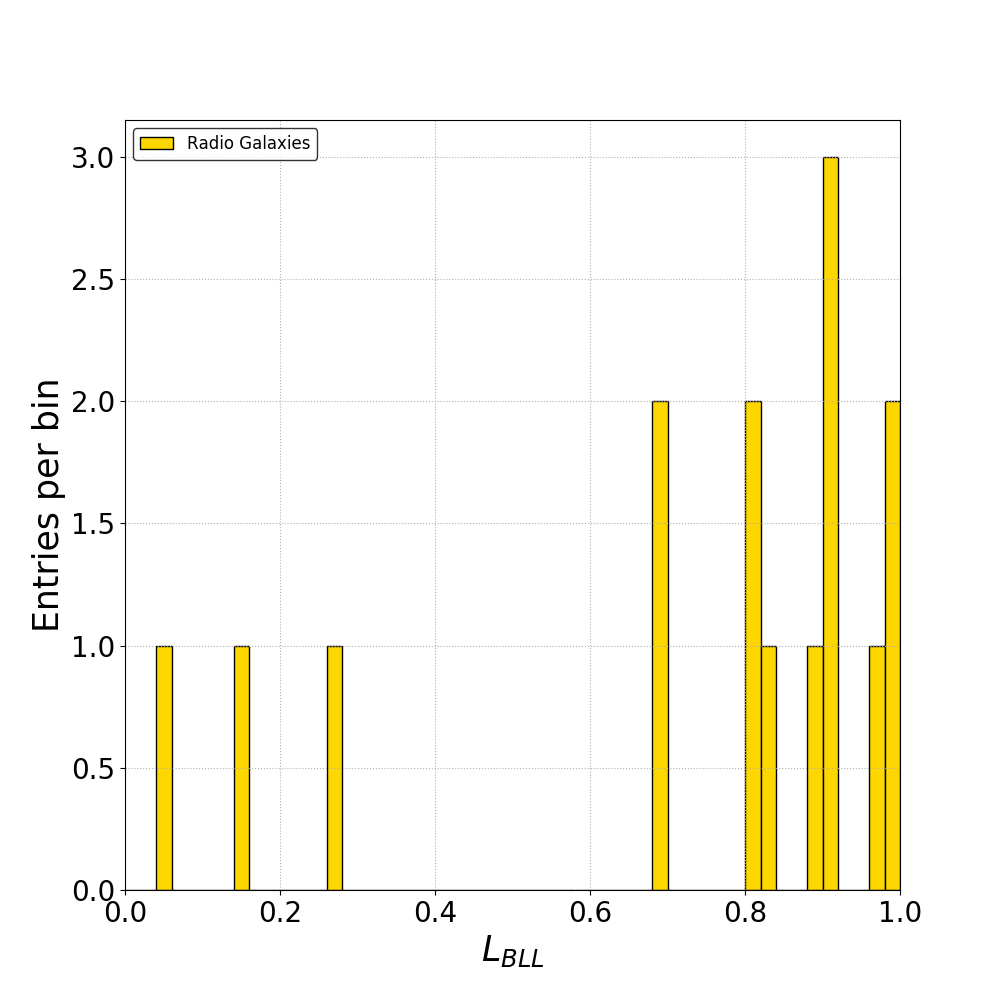}
\caption{(left plot): Distribution of the likelihood  to be a BL Lac candidate or FSRQ candidate \citep{bflap} applied to a testing sample of known  BL Lacs (blue) and FSRQs (red).  
(right plot):  Distribution of likelihood values using the same algorithm applied to  3FGL known $\gamma$-ray radio-galaxy MAGNs. }
\end{center}
\label{rdg}
\end{figure}

\begin{figure}
\begin{center}
\includegraphics[width=0.4 \textwidth]{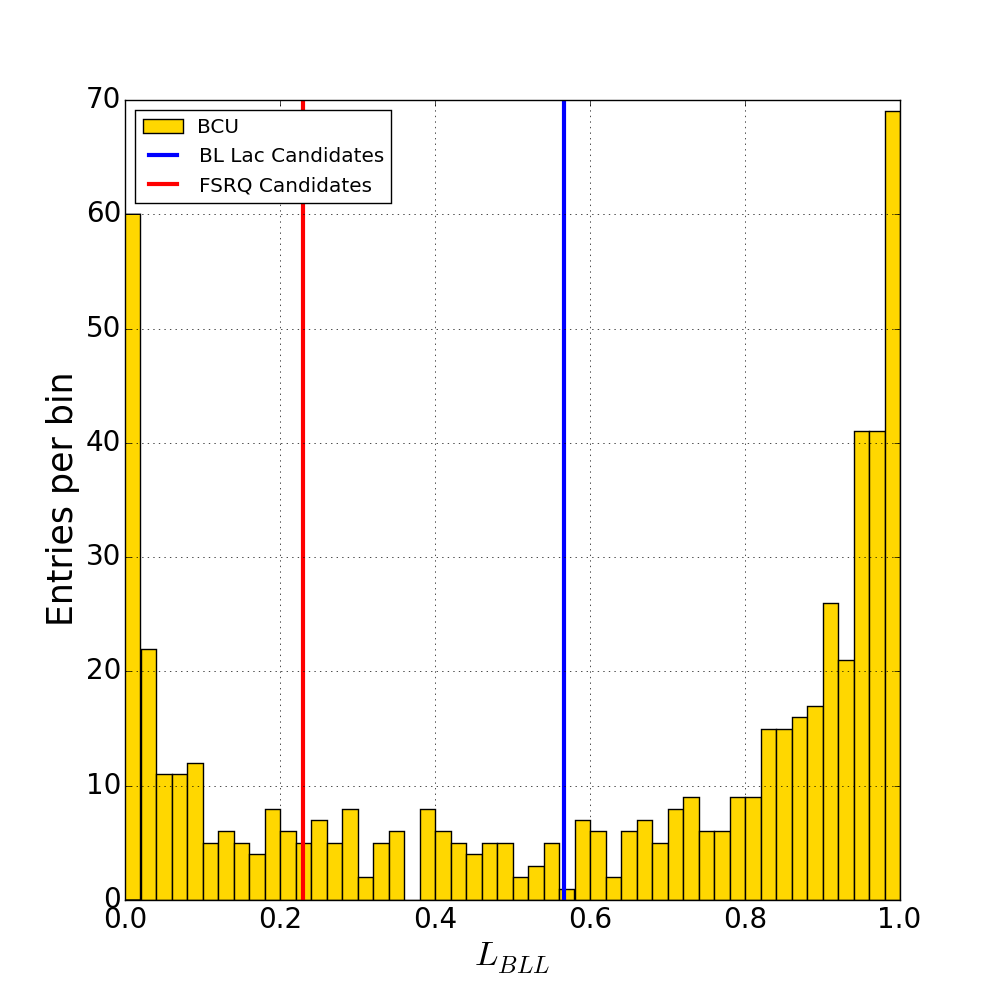}
\includegraphics[width=0.4 \textwidth]{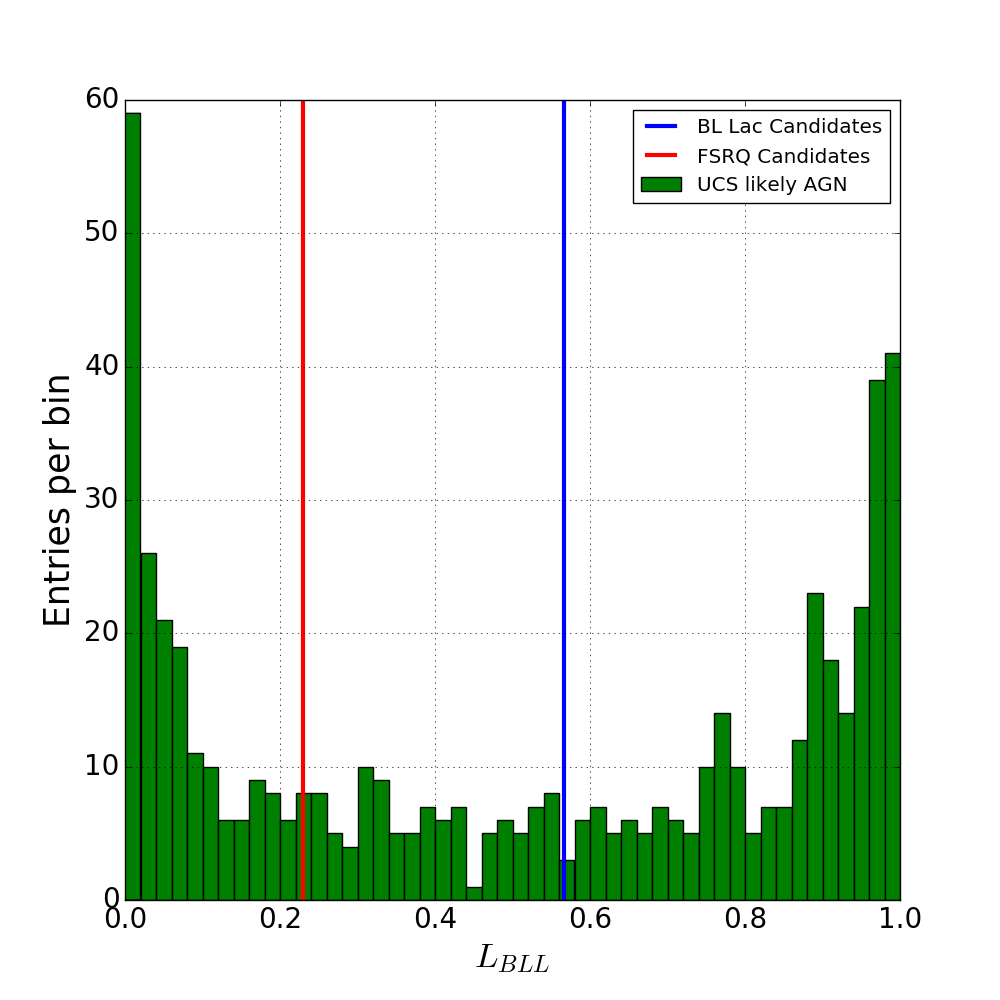}
\caption{Distribution of the likelihood of BCUs (left plot) and UCS$_{agn}$s (right plot) in the sample. Vertical blue and red lines indicate the classification thresholds of the algorithm as described in the text.} 
\end{center}
\label{ann}
\end{figure}

We further narrowed the screening using  the size of a galaxy as another  constraint to distinguish MAGN candidates from blazars, as following:

\begin{itemize}
\item We set 1 Mpc as the
maximum length of a  blazar jet.  This is a somewhat arbitrary choice. Although longer ones exist, those are exceptions. 
\item We assumed the maximum admissible jet orientation angle for a blazar to be 15\textdegree. 
\end{itemize}
Under these assumptions an approximate limit for the maximum projected length expected for a blazar jet is:
\begin{center}{1 Mpc $\times$ 0.5 $\times$ sin(15\textdegree) = 130 kpc.}         (1)
\end{center}

Sources with a projected size significantly bigger than 130 kpc are  more likely to be MAGNs than blazars.
Since the apparent size of an object is a function of the object distance from the observer, we calculated the size of our candidates in Table 2 whose redshift, from literature or by direct observations, was known.  
Four sources show a projected size exceeding 150 kpc, making them the \emph{best} MAGN candidates of this study.  All four came from the BCU set of candidates rather than from the UCS$_{agn}$ sources; however the small number of candidates  does not allow any firm conclusions about which types of 3FGL sources are most likely to be unrecognized MAGNs.

As another test, in Fig. 3 we used the comparison of $\gamma$-ray luminosity and spectral index as described in \citep{magn} \citet{paola} to highlight the positions of the four \emph{best} $\gamma$-ray MAGN candidates  (purple dots) within the 3FGL blazars and MAGN distribution. In the plot the candidates  are fully compatible with the 3FGL classified $\gamma$-ray MAGNs area (green dots). This comparison reinforces our initial hypothesis about the key features of a $\gamma$-ray MAGN candidate and the reliability of this study in finding some new candidates.  \\ 

\begin{figure*}
\begin{center}
\includegraphics[width=1.0 \textwidth]{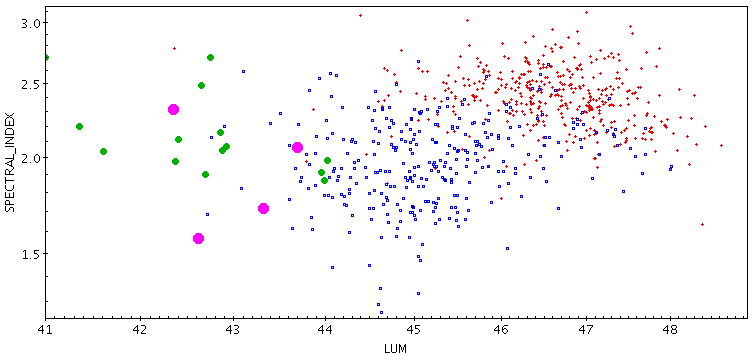}
\caption{Log($\gamma$-ray luminosity in erg s$^{-1}$) vs spectral index for 3FGL AGNs: FSRQ (red crosses), BL Lacs (blue squares), MAGNs (green dots), best $\gamma$-ray MAGN candidates (purple dots). Adopted from {\citep{paola}}.}
\end{center}
\label{4magn1}
\end{figure*}

\section {MAGN candidates}
\label{sec:RES}
In this section we discuss the $\gamma$-ray MAGN candidates, beginning with the four best sources highlighted with a diamond  in Table 2, along with other peculiar sources that deserve attention. When available we discuss the optical spectra  and other key parameters from the literature or obtained by our own observations. \\

\subsection{3FGL J0153.4+7114}
We confirm the conclusion of \citet{txs} that 3FGL J0153.4+7114 (TXS 0149+710, with z = 0.0239) is a low-power but core-dominated radio galaxy with 1.4 GHz radio flux density of 578 mJy.
The source was  also classified in the literature as a BL-Lac candidate \citep{otxs}, but this object shows strong asymmetry, with a wide jet directed in the W-NW direction, with kinks and bends. In the NVSS images \citep{0525} a single lobe structure prevails, with an extension of 193 kpc (Fig. 4). 
Although radio galaxies are often characterized by two lobes of extended emission, in some of them only one jet is visible, an effect referred to as Doppler favoritism \citep{ghis}.
 We collected an optical spectrum of TXS 0149+710 by direct observation at the 1.22 m telescope of Asiago Observatory, Italy. The source shows an optical spectrum (Fig. 5), with clear emission lines from H$\alpha$ and [N II] $\lambda 6584$, together with absorption lines from Mg I $\lambda 5175$ and Na I $\lambda 5893$. All lines show a redshift of $z = 0.0239$, consistent with the literature redshift of $z = 0.024$ \citep{otxs}. The continuum peaks at long wavelength, dropping towards the blue side of the spectrum, without signs of strong breaks or broad emission lines, suggesting the presence of only a faint or optically inefficient nuclear activity. The absence of broad spectral emission lines suggests that the source is not oriented along the line of sight. 

\subsection{3FGL J009.6$-$3211}
{3FGL J0009.6$-$3211 (IC 1531, with z= 0.0256) was already known in the 2FGL Catalogue as 2FGL J0009.9$-$3206 \citep{2fgl}. This source 
encompasses a radio source with 1.4 GHz radio flux density of 388.7 mJy and is characterized by an extension of $\sim$ 200 kpc in the southeast direction (Fig. 6) and a compact nuclear region.  IC 1531 was observed with the SAAO 1.9-m telescope using the SpUpNIC grating spectrograph on 2016 August 01 \citep{crause}. Observations were performed using grating 7 (300 lines/mm; 2.72 \AA~dispersion) with a total exposure time of 120 seconds. Data reduction was performed following the standard methods using the {\sc iraf/noao} packages. The spectrum (Fig. 7) shows the characteristic absorption lines of Ca II K\&H $\lambda$3933, $\lambda$3969, Mg $\lambda$5176 and Na I $\lambda$5893, commonly detected in elliptical galaxies. No emission lines are observed. The optical spectrum, therefore, indicates that the nuclear activity responsible for the radio and high-energy emission either lacks an intense ionizing radiation field or it lies beyond a deeply obscured region, oriented away from the direction of observation. 

\subsection{3FGL J0525.8$-$2014}
3FGL J0525.8$-$2014 (PMN J0525$-$2010 with z = 0.092) is considered as an elliptical galaxy with a 1.4 GHz radio flux density of 230.4 mJy. We obtained the  optical spectrum by direct observation at the 4.1-m Southern Astrophysical Research Telescope (SOAR) on 2017 January 29 using the Goodman High Throughput spectrograph. Slit width was 0.84 inches and the resolution is $\sim$ 830, with a dispersion of 1 $\AA$pixel$^{-1}$.  The spectrum (Fig. 8) shows a stellar continuum with clear absorption lines typical of an old stellar population, together with faint emission lines of H$\alpha$, [N~{\small II}] $\lambda6584$, [S~{\small II}] $\lambda\lambda6716,6731$ and [O~{\small III}] $\lambda\lambda4959,5007$. These emission lines, produced by an ionized gas component, may be suggestive either of star formation or of weak and obscured nuclear activity. The absence of any significant spectral contribution from a young stellar population and the high-energy activity of the source point towards AGN ionization, consistent with the presence of an active nucleus with an orientation axis lying away from the line of sight. The source has not been observed by the {\it Swift} X-ray Telescope and no other X-ray data are available; however the NVSS image in Fig. 9 shows two lobes with maximum radio extension of $\sim$ 430 kpc. The considerable angular extension excludes any blazar nature of the source. 

\subsection{3FGL J0039.0$-$2220} 
3FGL J0039.0$-$2220 (PMNJ0039$-$2220 with z = 0.064)  is a lenticular galaxy  with a radial velocity of 19301 km/s.  A lenticular galaxy is similar in content to an elliptical galaxy; they both have old stars. This class of galaxies lacks spiral arms due to the lack of star formation and looks like a very flat disc. The source was also listed in the 2FGL Catalogue (Nolan et al. 2012) as 2FGL J0038.7$-$2215. PMN J0039$-$2220 shows a 1.4 GHz radio flux density of 117.0 mJy. Fig. 10 shows the optical spectrum, available in the
6dFGRS DR3 Catalogue (Jones et al. 2009). The spectrum shows Ca H\&K, Na I, Mg I and [N II] spectral lines, which characterize the galaxy classification. Fig. 11 shows the NVSS image of a two-lobe extended structure with a maximum extension of $\sim$ 230 kpc. The presence of low ionization lines in the optical spectrum, combined with the absorption lines of an old stellar population, suggest the existence of a weak AGN activity, that, if seen under large angles, is consistent with the observed two-sided radio structure.

\section{Peculiar sources}
 Because of their features, two other peculiar sources among the 10 candidates are discussed in the following. Due to the lack of a complete set of data more analysis is needed in order to obtain a rigorous classification of these sources.

\subsection{3FGL J2204.4+0439}
3FGL J2204.4+0439 (4C +04.77 with z = 0.027)  is classified as BL Lac in the 3LAC Catalogue \citep{3lac} but as 5BZUJ2204+0440 in the BZCAT Catalogue \citep{BZCAT}, indicating a blazar of unknown type. This peculiar source shows an extended NVSS image (Fig. 12) with a size of $\sim$87 kpc, which is below the blazar jet size limit adopted as constraint  in this study, but a 1.4 GHz radio flux density of 468 mJy encourages more in-depth analysis. We studied the source by direct optical observation (Fig. 13), and the spectrum shows the characteristics of an elliptical galaxy, with absorption lines of Ca II K\&H, Mg I $\lambda 5175$ and Na I $\lambda 5893$ and a prominent 4000\AA~break. In addition, we detect the emission lines of H$\alpha$, together with faint contributions from [N II] $\lambda 6584$ and [S II] $\lambda 6731$, suggesting the presence of a low-ionization degree environment. The spectrum appears dominated by the host galaxy stellar population, suggesting that no radiatively efficient nuclear activity is operating in the source. 

\subsection{3FGL J0241.3+6542}
3FGL J0241.3+6542 (TXS 0237+655  with unknown redshift) has a 1.4 GHz radio flux density of 191 mJy. This source is very faint. We obtained an optical spectrum by direct observation at the 1.22 m telescope of Asiago Observatory, Italy (Fig. 14). No emission or absorption lines  were visible, so at this time we cannot do any estimation of redshift.  TXS 0237+655 is considered as a BCU but the source shows an extended / double NVSS structure seen in Fig. 15, which feeds suspicion towards a non-blazar source. More analysis and an optical spectrum by a large-mirror telescope are needed in order to classify this object. 

\section {Other  sources}
3FGL J0308.4$-$2852, 3FGL J0326.0$-$1842, 3FGL J0744.3+1715, 3FGL J1839.9+7646, 3FGL J2321.6$-$1619, our other candidates,  remain peculiar even if rigorous analysis has not been possible because few data at other frequencies were available. 
3FGL J0308.4$-$2852, an unassociated AGN in 3FGL, is the only source that shows a NVSS double-like  structure (Fig. 16) but without an optical spectrum its classification remains uncertain. 3FGL J0744.3+1715 and 3FGL J2321.6$-$1619 have been observed with the Very Large Array and the Australia Telescope Compact Array \citep{franc} in a radio observation campaign of $\gamma$-ray  unassociated sources and are identified as extended sources.

\subsection{TeV MAGNs candidates} 

MAGNs are faint sources at GeV energies and consequently rare in the TeV sky \citep[e.g.,][]{angio, ele}. 
Because two of the MAGN candidates  considered in this study,  TXS 0149+710 and TXS 0237+655, are also in the Third Catalogue of Hard Fermi LAT sources \citep[3FHL][]{3fhl}, which reported $\gamma$-ray sources detected above 10 GeV with the {\it Fermi} LAT, those sources could be TeV-energy candidates too. 
We therefore compare the extrapolated fluxes of these sources against the sensitivity of present IACTs and the future Cherenkov Telescope Array (CTA).  We adopted the {\it Fermi}-LAT spectral shape of the sources in the range between 0.1 and 300 GeV using the best-fit model parameters from
the LAT 3FGL Catalogue\footnote{The file can be retrieved from the \emph{Fermi} Science Support Center: \url{http://fermi.gsfc.nasa.gov/ssc/data/access/lat/4yr\_catalog/}.} and particularly we referred to the following relation derived from the spectral model that fits the data:

\begin{equation}
F {E^2} = n_0 \left ( \frac{E}{E_0} \right )^{-\alpha} {E^2} 
\end{equation}

where $n_0$ is a prefactor,  $\alpha$ is the spectral index, ${E_0}$ is the folding energy (1 TeV)  and  the energy flux (F${E^2}$)  is expressed in units of erg cm$^{-2}$ s$^{-1}$.\\
In order to realistically assess whether the  two sources are good candidates for VHE observations,  $\gamma$-ray absorption on the extragalactic background light (EBL) 
has to be taken into account.  The EBL spans the wavelength regime between ultraviolet and far-infrared wavelengths and mainly consists of the integrated starlight emitted over the history of the Universe and starlight  absorbed and re-emitted by dust in galaxies \citep{hauser,kashlinsky}.  The EBL photons interact with $\gamma$~rays to produce electron-positron pairs, which process attenuates the initial $\gamma$-ray flux \citep{nikisov,gould,dwek}. The exponential attenuation monotonically increases with blazar distance  and $\gamma$-ray energy. 
We show the effect of the attenuation in Fig. 17, where we show the BCU spectra as measured in the 3FGL and 3FHL, the latter extending to energies of 2\,TeV. 
As only the redshift of TXS 0149+710  is known ($z = 0.0239 $), we show the EBL attenuation for both sources at redshifts of $z = 0.02 $  and $z = 0.1 $ using the EBL model of \citet{dominguez}.
Such a redshift is compatible with the observations in the 3FHL for all sources except 3FGL\,J1741.1$-$2029, where the data point at $\sim300\,$GeV indicates a lower redshift. 
We show the CTA sensitivity for 50\,hours (5\,hours) of observations as a solid (dashed) grey line in Fig. 17.
The CTA sensitivity for 5\,hours of observations is similar to that of currently operating IACTs for 50\,hours of observations except with a higher threshold energy of $\sim 80\,$GeV. \footnote{The sensitivity curves of the northern and southern array  are available at \url{www.cta-observatory.org}}
The simulation results show that TXS 0149+710 is potentially detectable within 50\,hours of CTA observations but is unlikely to be seen with current IACTs.  TXS 0237+655, whose redshift is not available but has a spectral index harder than that of TXS 0149+710,  might be detectable by currently operating IACTs and appears to be a good candidate for detection with CTA.

\section{Conclusion}
\label{sec:CON}
Motivated by wishing to develop a process to expand the small sample of $\gamma$-ray MAGN, we have developed a search method based on machine learning combined with comparisons to parameters of the known $\gamma$-ray MAGN.  The 10 identified candidates represent in many ways a first step toward that goal.  Follow-up work is necessary to firmly establish these sources as MAGNs, including finding redshifts for half the candidates, using deeper radio, optical, and X-ray observations to reveal morphological details, and possibly undertaking Very Long Baseline Interferometry observations to examine the jets themselves. We also recognize that this search is necessarily incomplete. As already pointed out by \citet{3lac}, the $\gamma$-ray sources with unknown properties are generally fainter than the well-defined classes.  The fainter sources offer less of the variability information required for the machine learning method, and so there may be $\gamma$-ray MAGN among the 529 3FGL sources eliminated in the first step of our method. The level of incompleteness is difficult to quantify with such a small number of sources.  Nevertheless, the best candidates among our set are convincing as $\gamma$-ray MAGN and can be combined with the known 3FGL MAGN for population studies.  Two of our candidates are also promising targets for TeV observations.  The success of the method also indicates that it will be useful in searching for more $\gamma$-ray MAGN in the forthcoming 4FGL catalog, particularly when combined with the ongoing Very Large Array Sky Survey (VLASS)\footnote{\url{https://science.nrao.edu/science/surveys/vlass/}}.

\section{Acknowledgments}
Support for science analysis during the operations phase is gratefully acknowledged from the {\it Fermi}-LAT Collaboration for making the 3FGL results available in such a useful form. This study is based on  observations collected at the 1.22 m Copernico telescope (Asiago, Italy) of the INAF - Osservatorio Astronomico di Padova, on observations obtained at the South African Astronomical Observatory (SAAO), with the research supported in part by the National Research Foundation of South Africa for the grant No. 87919, and on observations obtained at the Southern Astrophysical Research (SOAR) telescope,  which is a joint project of the Minist\'{e}rio da Ci\^{e}ncia, Tecnologia, e Inova\c{c}\~{a}o (MCTI)  da Rep\'{u}blica Federativa do Brasil, the U.S. National Optical Astronomy Observatory (NOAO),  the University of North Carolina at Chapel Hill (UNC), and Michigan State University (MSU). We also particularly thank Paola Grandi of the Institute of Space Astrophysics and Cosmic Physics of Bologna - Italy (IASF/INAF)  for her precious comments.

\begin{figure}
\begin{center}
\includegraphics[width=0.5 \textwidth]{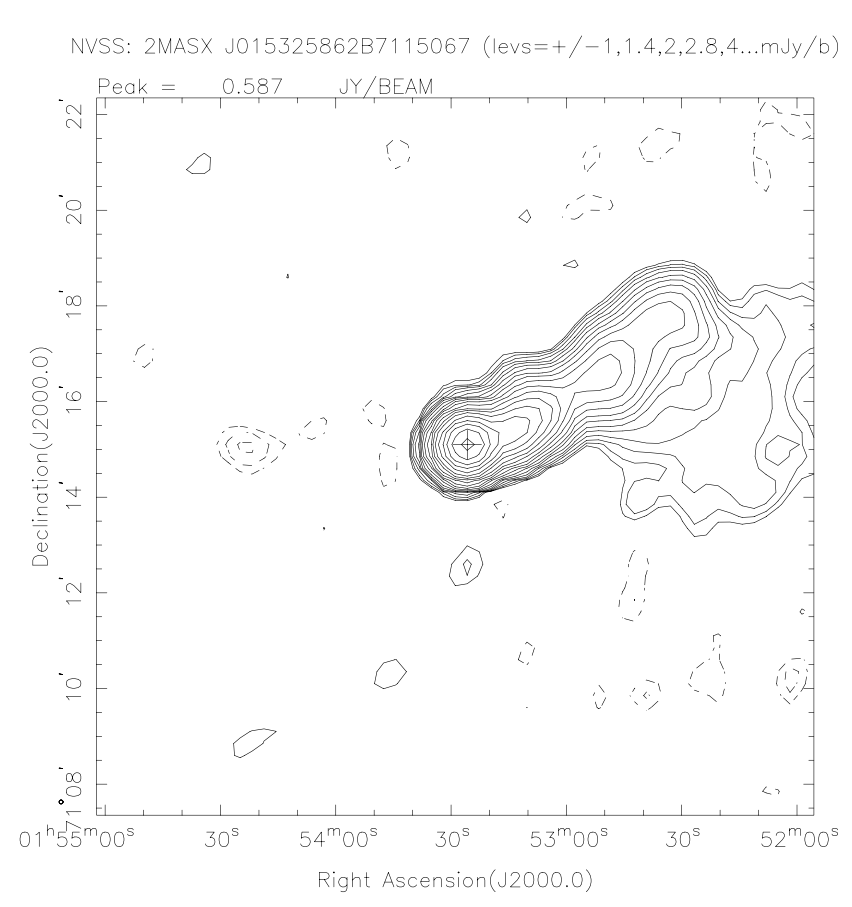}
\caption{TXS 0149+710 NRAO VLA Sky Survey (NVSS) image at 1.4 GHz. where a single lobe structure prevails, with an extension of 173 kpc.}
\end{center}
\label{txsOTT}
\end{figure}

\begin{figure}
\begin{center}
\includegraphics[width=0.5 \textwidth]{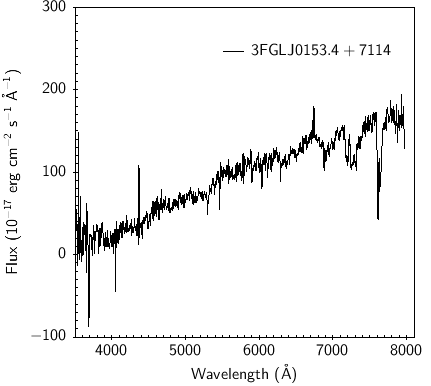}
\caption{TXS 0149+710 optical spectrum \citep{otxs}. The presence of narrow emission lines, together with the absorption lines and continuum of an old stellar population, indicates the existence of an ionized gas component that might originate from the central high-energy source. The absence of broad spectral emission lines suggests that the source is not oriented along the line of sight and its central regions are therefore not visible.}
\end{center}
\label{txsOTT}
\end{figure}

\begin{figure}
\begin{center}
\includegraphics[width=0.5 \textwidth]{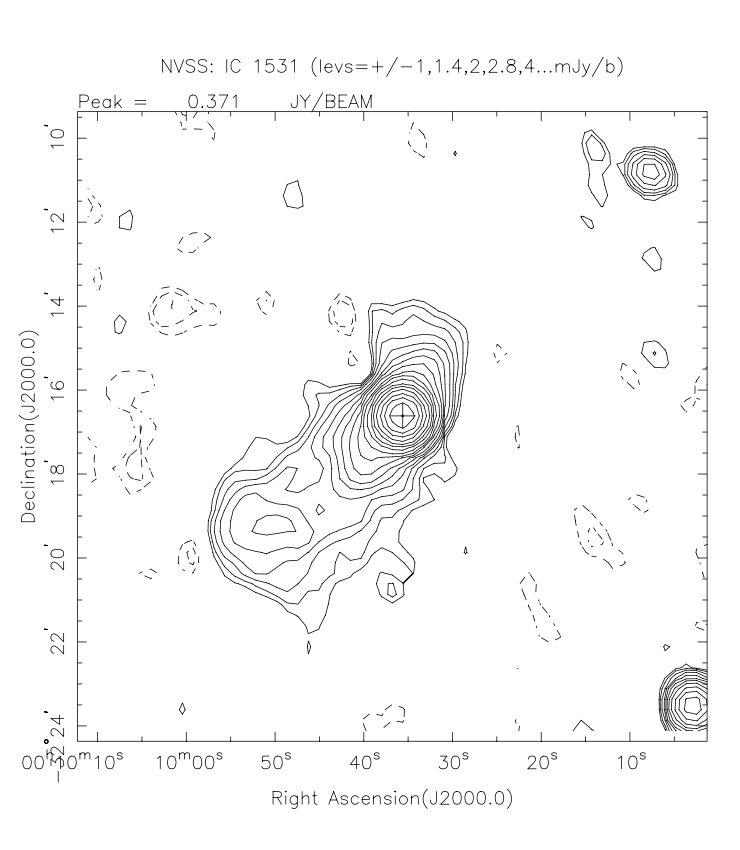}
\caption{NRAO VLA Sky Survey (NVSS) image at 1.4 GHz image of IC1531 characterized by the presence of an extension of $\sim$ 200 kpc in the southeast direction.}
\end{center}
\label{ic1531nvss}
\end{figure}

\begin{figure}
\begin{center}
\includegraphics[width=0.5 \textwidth]{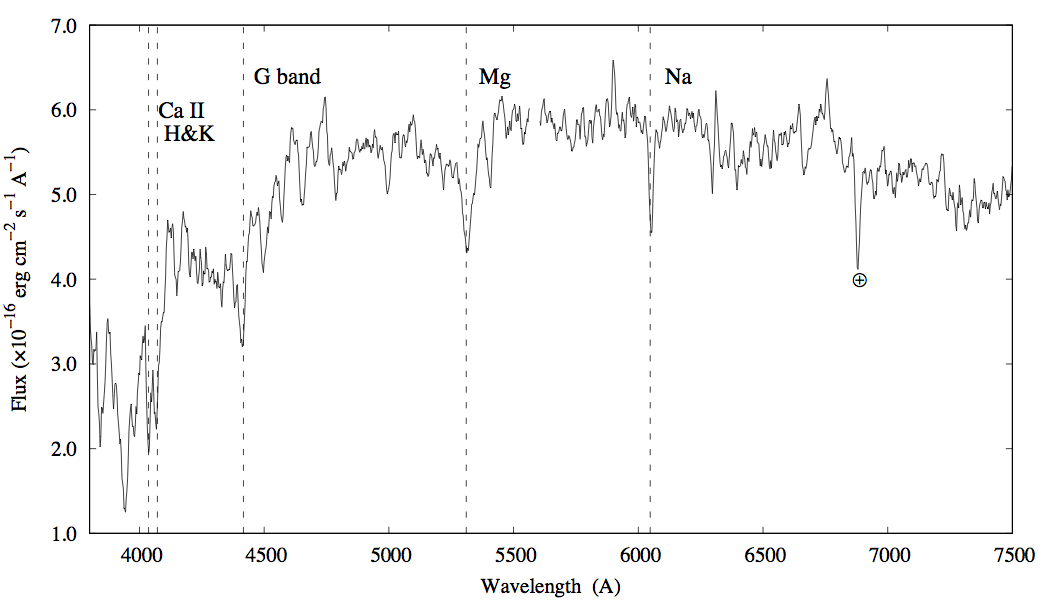}
\caption{The optical spectrum of IC 1531 shows the absorption lines and continuum of an elliptical galaxy, with no signs of ionized gas. This suggests that the origin of the radio and high-energy signals lies in a heavily obscured region, very likely oriented far away from the line of sight. }
\end{center}
\label{ic1531ott}
\end{figure}

\begin{figure}
\begin{center}
\includegraphics[width=0.5 \textwidth]{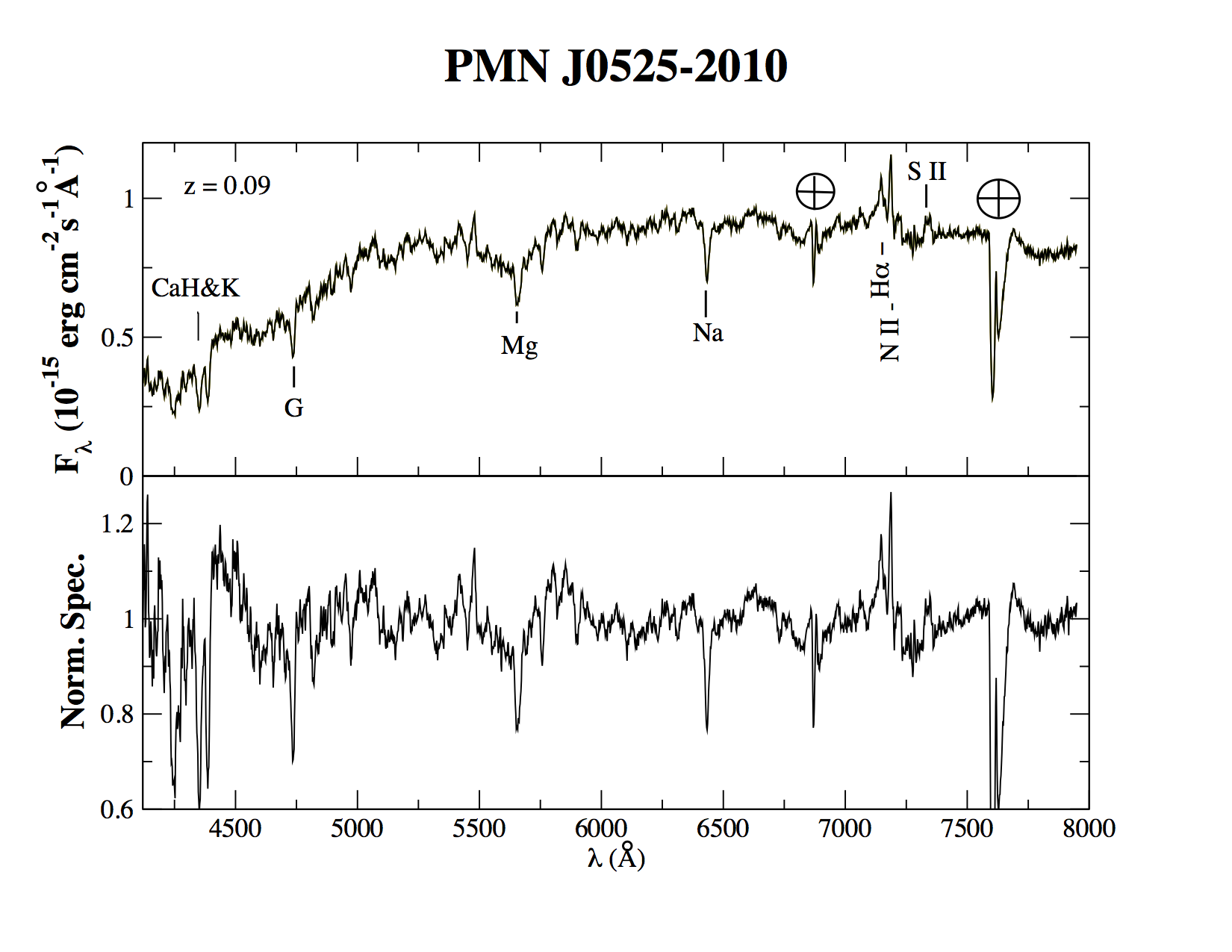}
\caption{The optical spectrum of PMN J0525$-$2010 shows faint emission lines together with the characteristic absorption lines and continuum of an aged stellar population. The spectrum is therefore consistent with the presence of gas ionized by an active nucleus oriented away from the line of sight.}
\end{center}
\label{pmn525nvss}
\end{figure}

\begin{figure}
\begin{center}
\includegraphics[width=0.5 \textwidth]{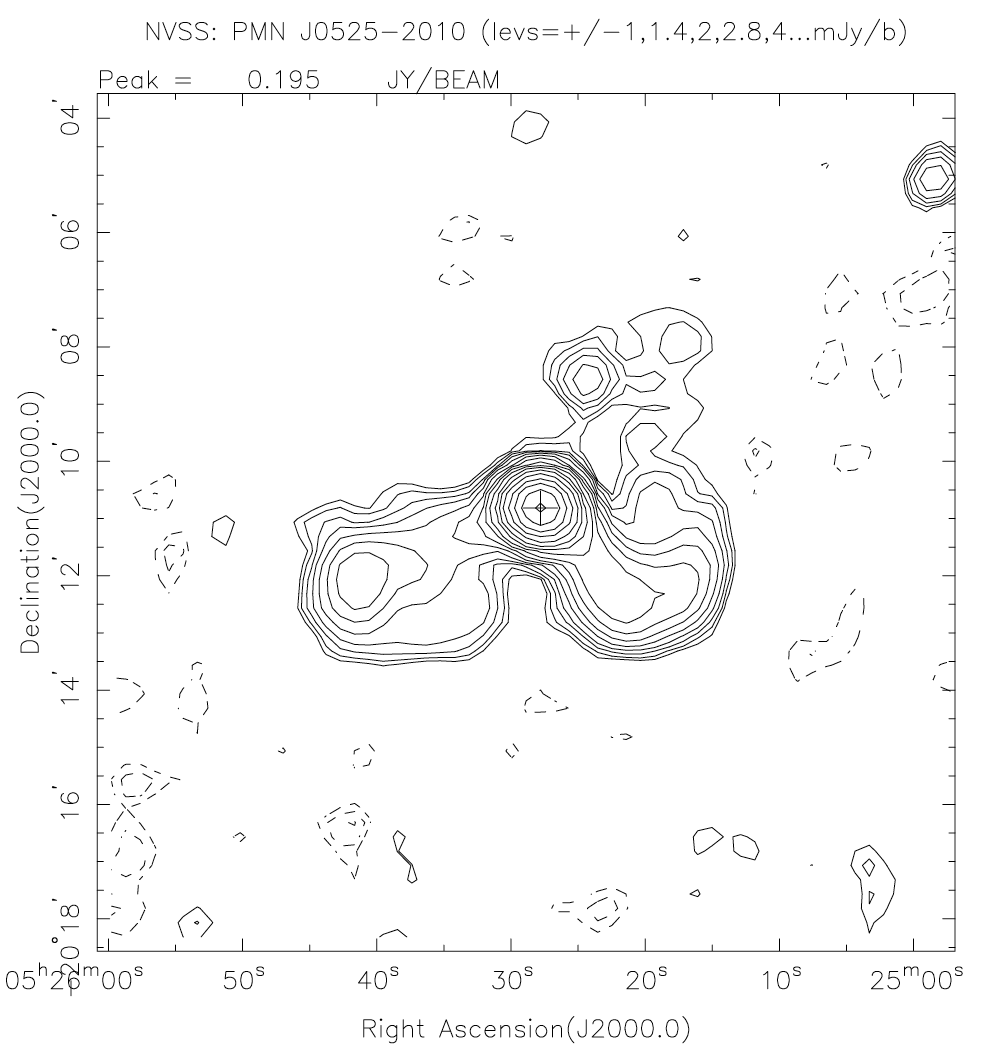}
\caption{NRAO VLA Sky Survey (NVSS) image at 1.4 GHz of PMNJ0525.8$-$2010 at 1.4 GHz, which shows a two-lobe structure with maximum extension of $\sim$ 430 kpc. }
\end{center}
\label{pmn525nvss}
\end{figure}

\begin{figure}
\begin{center}
\includegraphics[width=0.5 \textwidth]{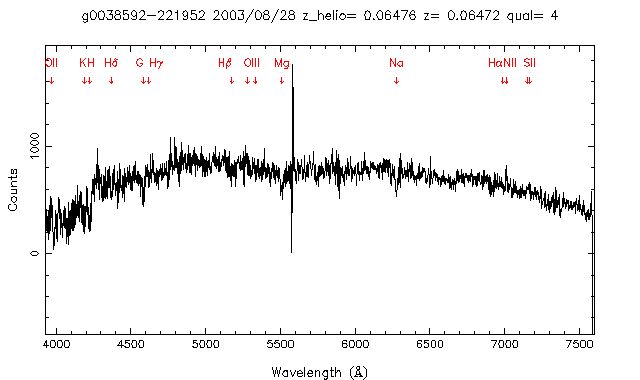}
\caption{PMNJ0039$-$2219 optical spectrum from the 6dFGRS-DR3 Catalogue \citep{6df}.  Ca H$\&$K , Na I, Mg I and [N II] spectral lines characterize the galaxy classification.}
\end{center}
\label{pmn39ott}
\end{figure}

\begin{figure}
\begin{center}
\includegraphics[width=0.5 \textwidth]{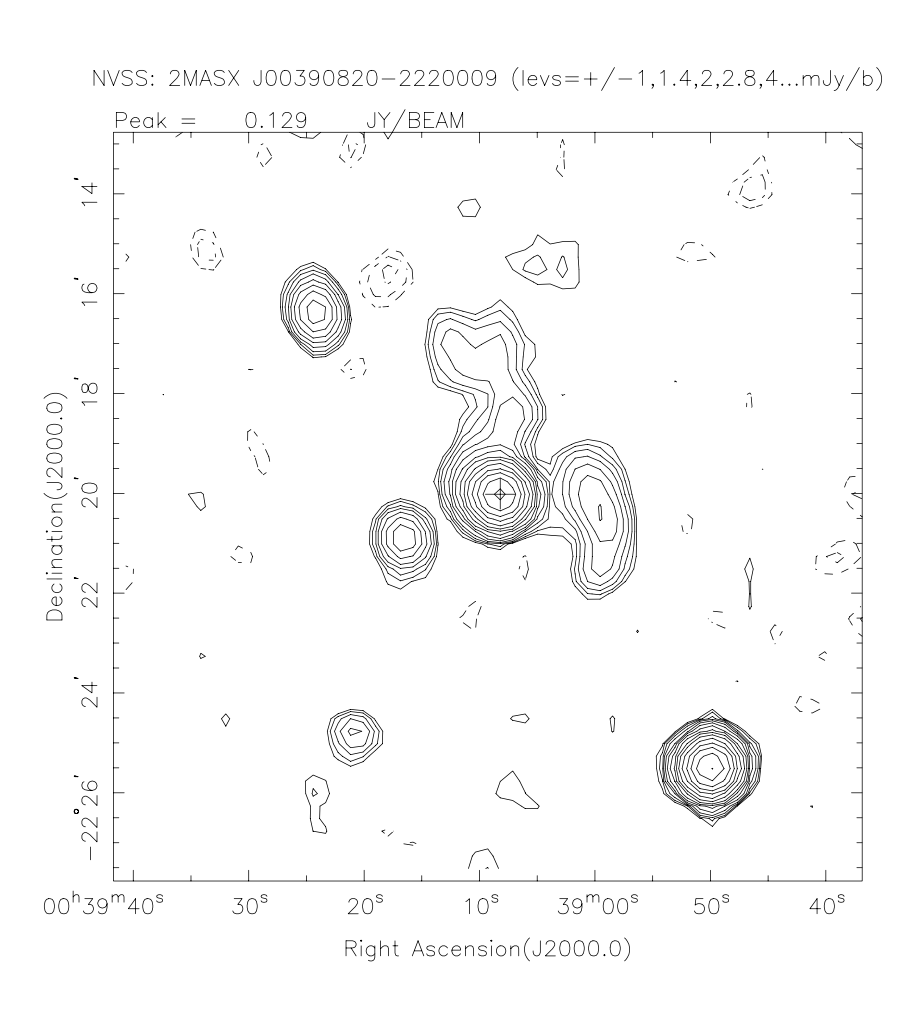}
\caption{NVSS image of of 3FGL J0039.0$-$2218 / PMN J0039$-$2220 at 1.4 GHz}
\end{center}
\label{pmn39nvss}
\end{figure}

\begin{figure}
\begin{center}
\includegraphics[width=0.5 \textwidth]{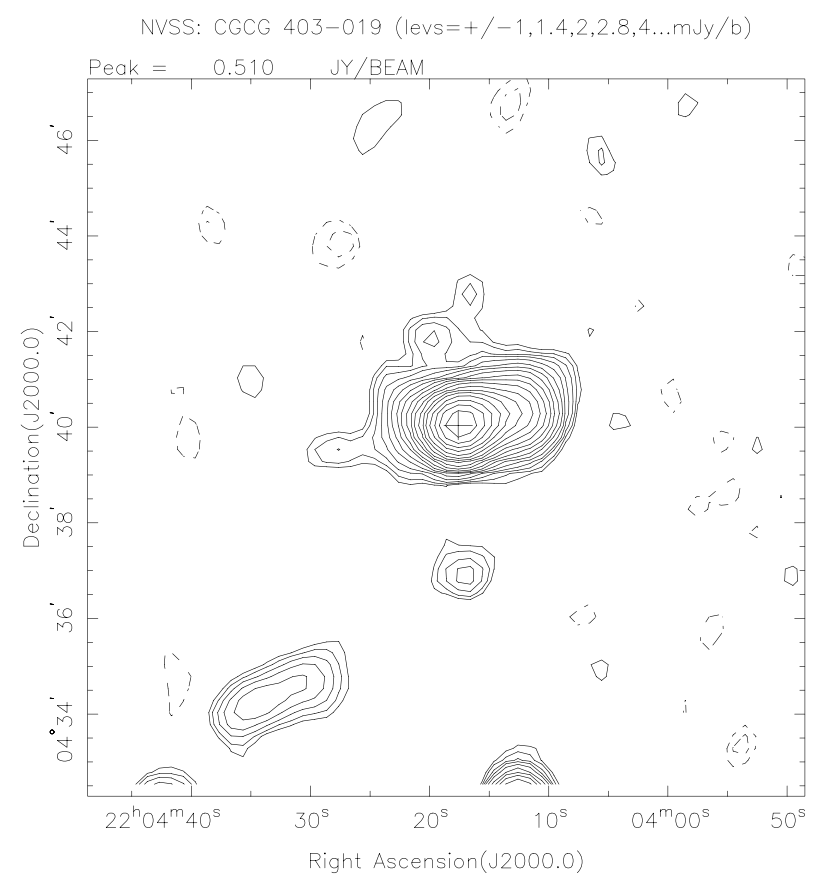}
\caption{NRAO VLA Sky Survey (NVSS) image at 1.4 GHz of 4C +04.77. The source shows an extended structure characterized by an extension of $\sim$ 87 kpc. }
\end{center}
\label{4cnvss}
\end{figure}

\begin{figure}
\begin{center}
\includegraphics[width=0.5 \textwidth]{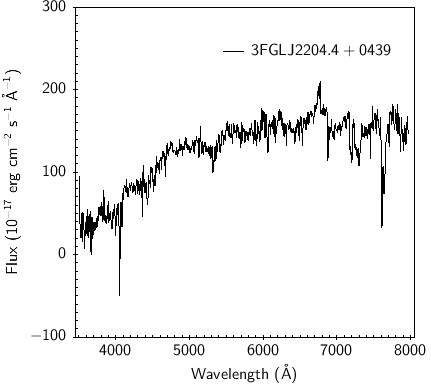}
\caption{Optical spectrum of 4C +04.77. The spectrum appears dominated by the host galaxy stellar population, with just a faint H$\alpha$\ emission line, suggesting that no radiatively efficient nuclear
activity is operating in the source.}
\end{center}
\label{4cottico}
\end{figure}

\begin{figure}
\begin{center}
\includegraphics[width=0.5 \textwidth]{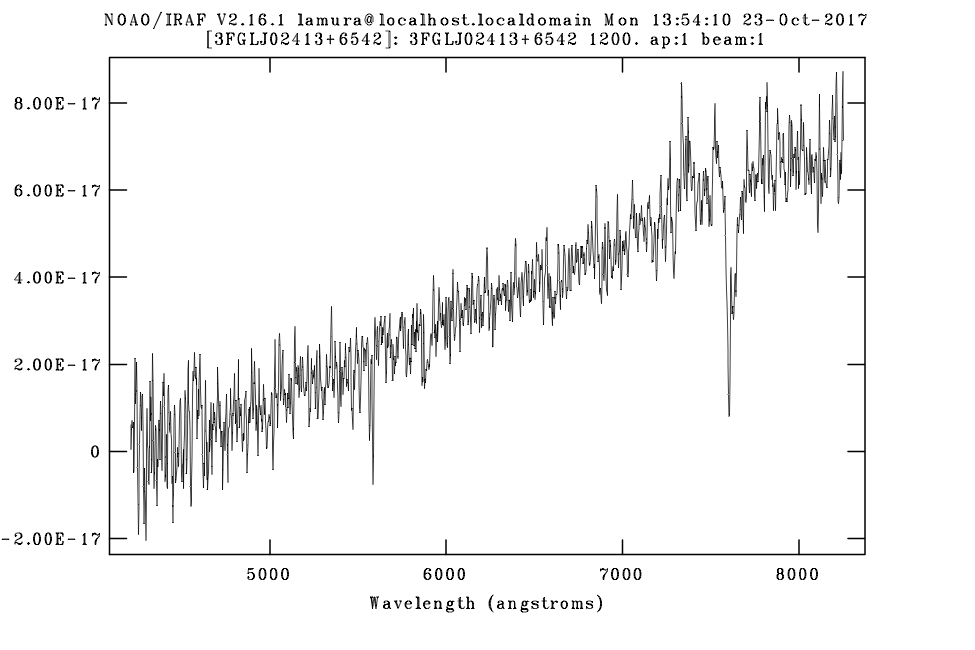}
\caption{Optical spectrum of TXS 0237+655. No emission lines  are visible, so at this time no affirmation on redshift  is available.}
\end{center}
\label{4cnvss}
\end{figure}

\begin{figure}
\begin{center}
\includegraphics[width=0.5 \textwidth]{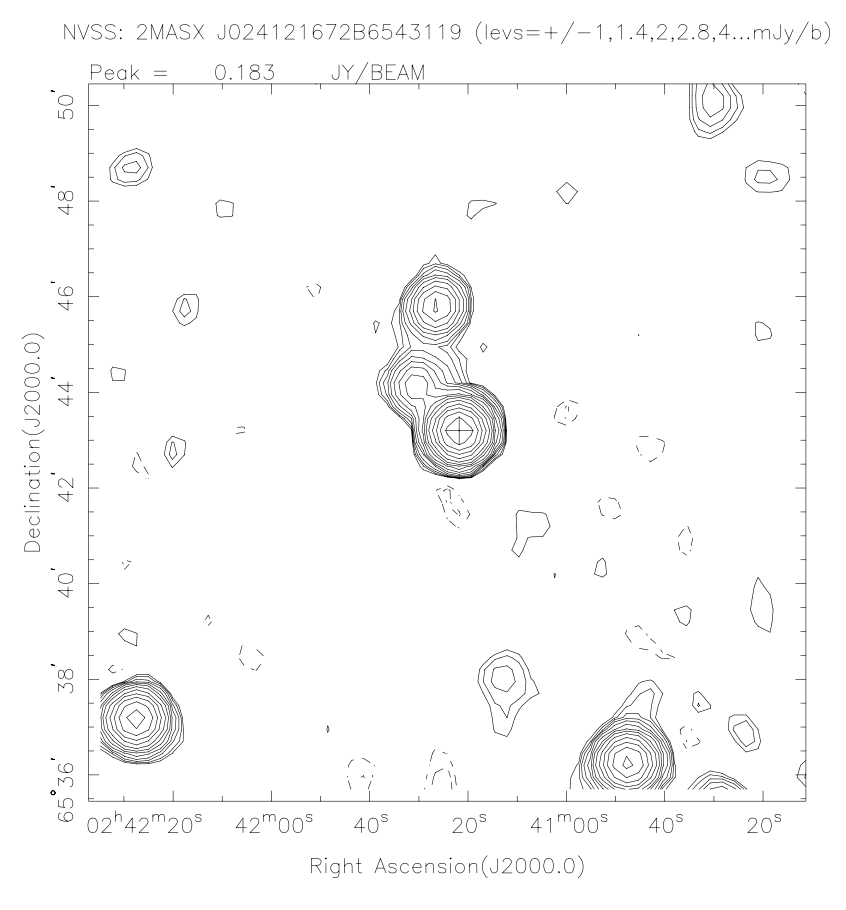}
\caption{NRAO VLA Sky Survey (NVSS) image at 1.4 GHz of TXS 0237+655, which shows a double extended structure that could represent a MAGN candidate.}
\end{center}
\label{4cnvss}
\end{figure}


\begin{figure}
\begin{center}
\includegraphics[width=1.1 \textwidth]{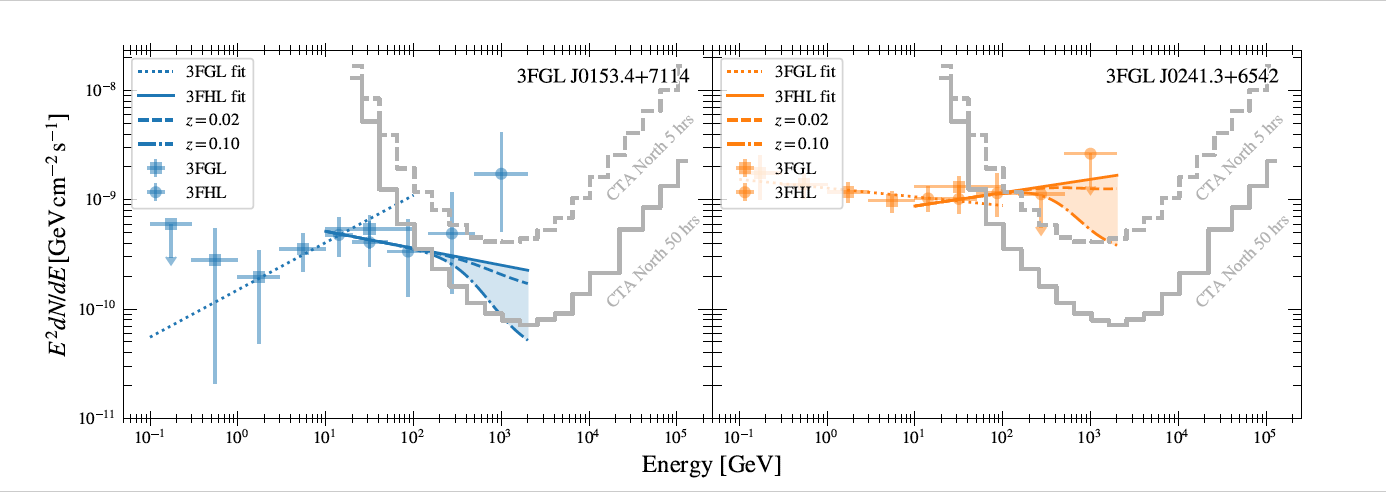}
\caption{Detectability of TXS 0149+710 and TXS 0237+655  in the energy range 100 GeV $-$ 2 TeV, based on the 3FGL and 3FHL fits. The dashed line represent the simulation at redshift z=0.02 and the dot - dashed line at redshift = 0.1.  The source  TXS 0149+710 shows a possible detectability at $E_{r}$ = 1 TeV for 50 hours of CTA exposure. TXS 0237+655  also shows a possible detectability by the present generation of IACTs if we consider the IACTs sensitivity $\sim$ CTA 5h. }
\end{center}
\end{figure}

\begin{table}
\begin{center}
\caption{MAGN candidates selected from sources with uncertain classification (UCS$_{agn}$: unassociated; BCU: blazar of uncertain type; BL Lac) in the 3FGL {\it Fermi}-LAT Catalogue. P.I. and V.I. are the power-law index and the variability index of the $\gamma$-ray source, from 3FGL. The four best candidates are preceded by a diamond. }\begin{tabular}{lcccccccccccccccr}
\hline
\hline
\bf{3FGLname}&\bf{Assoc.name} &\bf{3FGL class}& \bf{RA} &\bf{Dec} &\bf{P.I.} &\bf{V.I.} &{\bf{Size}} kpc\\
\hline\hline
$\diamond$ 3FGL J0009.6$-$3211&IC 1531&BCU&00 09 36.3&-32 11 55.28&2.31&42.5&200\\
$\diamond$ 3FGL J0039.0$-$2218&PMN J0039$-$2220&BCU&00 39 00.1&-22 18 24.4&1.71&36.5&290\\
$\diamond$ 3FGL J0153.4+7114&TXS 0149+710&BCU&01 53 26&+71 14 30&1.56&36.6&193\\
3FGL J0241.3+6542	&TXS 0237+655&UCS$_{agn}$&02 41 21.79&	+65 43 11.4&2.08&33.4&&\\
3FGL J0308.4$-$2852&&UCS$_{agn}$&03 08 16.94&-28 51 03.8&2.53&26.1\\
3FGL J0326.0$-$1842&PMN J0325$-$1843&BCU&03 25 54.99&-18 44 08.8&2.20&43.2\\
$\diamond$ 3FGL J0525.8$-$2014&PMN J0525$-$2010&	BCU	&05 25 48.19&	-20 14 53.8&	2.06&	51.1& 431\\
3FGL J2204.4+0439&	4C +04.77&	BL Lac	&22 04 24&	+04 39 56&	2.12&	59.8 & 87 \\
3FGLJ0744.3+1715	&&	UCS$_{agn}$	&07 44 12	&+17 17 48	&2.26&	31.5\\
3FGLJ1839.9+7646&&		UCS$_{agn}$&	18 41 10&	+76 48 41&	1.92&	50.1\\
 \hline
\end{tabular}
\end{center}
\end{table}

\label{lastpage}

\end{document}